# METAL TO INSULATOR TRANSITION IN THE 2-D HUBBARD MODEL: A SLAVE BOSON APPROACH

Raymond Frésard and Klaus Doll*

Institut für Theorie der Kondensierten Materie
Universität Karlsruhe
Postfach 6980
D-76128 Karlsruhe

## INTRODUCTION

Since the discovery of the High-$T_c$ superconductors, [1], the Hubbard model has been the subject of intense investigations following Anderson's proposal [2] that the model should capture the essential physics of the cuprate superconductors. From the earlier attempts to obtain the magnetic phase diagram on the square lattice (for an overview see the book by Mattis [3]) one can deduce that antiferromagnetic order exists in the vicinity of the half-filled band whereas ferromagnetic ordering might take place in the phase diagram for strong repulsive interaction strength and moderate hole doping of the half-filled band. Obviously antiferromagnetic and ferromagnetic orders compete in this part of the phase diagram. More recent calculations [4] established that the ground state of the Hubbard model on the square lattice shows long-ranged antiferromagnetic ordering with a charge transfer gap. However, the problem of mobile holes in an antiferromagnetic background remains mostly unsolved. Suggestions for a very wide ferromagnetic domain in the phase diagram based on the restricted Hartree-Fock Approximation have been made by several authors [5] on the cubic lattice, and on the square lattice [6–8]. This domain appears for large interaction and moderate hole doping in which case the Hartree-Fock Approximation ceases to be controlled. Within this framework one expects to obtain reliable results for moderate $U$ where the paramagnetic phase is indeed unstable towards an incommensurate spin structure at a critical density $n_c(U)$ [9]. The Gutzwiller Approximation (GA) [10–12] has been applied [13], even for large $U$, yielding results similar to the Hartree-Fock Approximation. However, for large $U$, a ferromagnetic domain appears only if the density is larger than some critical value. In the Kotliar and Ruckenstein slave boson technique [14] the GA appears as a saddle-point approximation of this field theoretical representation of the Hubbard model. In the latter a metal-insulator transition occurs at half-filling as recently discussed by Lavagna [15]. The contribution of the thermal fluctuations has been calculated [16] and turned out to be incomplete as this representation, even though exact, is not manifestly spin-rotation invariant. Spin-rotation invariant [17] and spin and charge-rotation invariant [18] formulations have been proposed and the first one was used to calculate correlation functions [19] and spin fluctuation contributions to the specific heat [20]. Comparisons of ground state energy with Quantum Monte-Carlo simulations, including antiferromagnetic ordering [21] and spiral states [22], or with exact diagonalisation data [23] have been done and yield excellent agreement, and a magnetic phase diagram has been proposed [24].

Such calculations are not restricted to the square lattice and can be performed, for example, on the honeycomb lattice, where a (semi-) metal to insulator transition takes place at half-filling when the interaction strength is increased [25]. In this paper we first compare the slave-boson estimate of the critical interaction strength with the Quantum Monte-Carlo simulations of Sorella and Tossati [25]. In a second part we determine the phase boundary of the paramagnetic state on the square lattice by determining the instability line of the latter with respect to an arbitrary long-ranged incommensurate spin order at finite temperature, thus extending the calculations we performed at zero temperature [26].

## HONEYCOMB LATTICE RESULTS

Considering the honeycomb (HC) lattice instead of the square lattice allows for a further study of the interaction induced Metal-to-Insulator Transition (MIT). It has been shown by Quantum Monte-Carlo simulations [25] that, at half-filling, the system undergoes a MIT as the interaction strength is raised up. This follows from the fact that we meet here a semi-metal in which, as far as we consider the Stoner criterion, no long-ranged magnetic order is expected for small $U$. On the other hand the HC lattice is bipartite and can obviously support un-frustrated anti-ferromagnetism at large $U$. The two considerations thus yield a scenario for the MIT at half-filling. By raising up the interaction long-ranged anti-ferromagnetic order will set in and a gap will open at a critical interaction strength, thus yielding a transition from a metallic to an insulator behavior. Another possibility is provided by the Brinkman-Rice mechanism. Here raising up the interaction leads to a diverging effective mass [11, 12] at a critical value of $U$, and a gap opens in the charge excitation spectrum [15, 27, 28] at the Fermi energy. It thus describes a Mott insulator. These two scenarii can be addressed in the slave boson mean-field approach.

On the HC lattice, the free electron dispersion relation is given by:

$$t_{\vec{k},\nu} = 2t\nu \mid 1 + \exp\left(-i\vec{k} \cdot \vec{b}_1\right) + \exp\left(-i\vec{k} \cdot \vec{b}_2\right) \mid \qquad \nu = \pm 1 \qquad (1)$$

where, $a$ being the lattice constant, $\vec{b}_1 = (3, \sqrt{3})a/2$, $\vec{b}_2 = (3, -\sqrt{3})a/2$ and the pseudo-fermions acquires the effective dispersion relation $E_{\vec{k},\sigma,\nu} = z_0^2 t_{\vec{k},\nu} + \beta_0 - \mu_0$ where $z_0^2$ denotes the inverse effective mass of the quasi-particles:

$$z_0^2 = \frac{1}{1 - d^2 - \frac{p_0^2}{2}} \frac{p_0^2}{2}(e+d)^2 \frac{1}{1 - e^2 - \frac{p_0^2}{2}} \qquad . \qquad (2)$$

The paramagnetic mean-field energy reads:

$$F = -T \sum_{\vec{k},\sigma,\nu} \ln\left(1 + e^{-(E_{\vec{k},\sigma,\nu}/T)}\right) + Ud^2 + \alpha(e^2 + p_0^2 + d^2 - 1) - \beta_0(2d^2 + p_0^2) \qquad (3)$$

where $e^2, p_0^2, d^2$ denote respectively the average occupation number of empty, singly-occupied, and doubly-occupied sites ; $\alpha$ and $\beta_0$ are the Lagrange multipliers enforcing the constraints [18]. When staggered order is allowed, the effective dispersion relation for the pseudo-fermions becomes:

$$E_{\vec{k},\sigma,\nu} = \beta_0 - \mu_0 + \sigma\sqrt{\beta^2 + z_\sigma^2 z_{-\sigma}^2 t_{\vec{k},\nu}^2} \qquad \sigma = \pm 1 \qquad (4)$$

where

$$z_\sigma = \frac{1}{\sqrt{1 - d^2 - \frac{(p_0+\sigma p)^2}{2}}} \frac{p_0(e+d) + \sigma p(e-d)}{\sqrt{2}} \frac{1}{\sqrt{1 - e^2 - \frac{(p_0-\sigma p)^2}{2}}} \qquad (5)$$

and the free energy reads:

$$F = -T \sum_{\vec{k},\sigma,\nu} \ln\left(1 + e^{-(E_{\vec{k},\sigma,\nu}/T)}\right) + Ud^2 + \alpha(e^2 + p_0^2 + p^2 + d^2 - 1) - \beta_0(2d^2 + p_0^2 + p^2) - 2p_0\vec{\beta}\cdot\vec{p}$$
$$(6)$$

We minimize both free energies eq. (3), eq. (6) at zero temperature. The resulting free energies are displayed in fig. 1a. The agreement with the Quantum Monte-Carlo values of Sorella and Tossati [25] is excellent as long as no magnetic order occurs, even though no quantum fluctuations are included in the calculation. The differences range from 5% at $U = 4t$ to 11% at $U = 8t$.

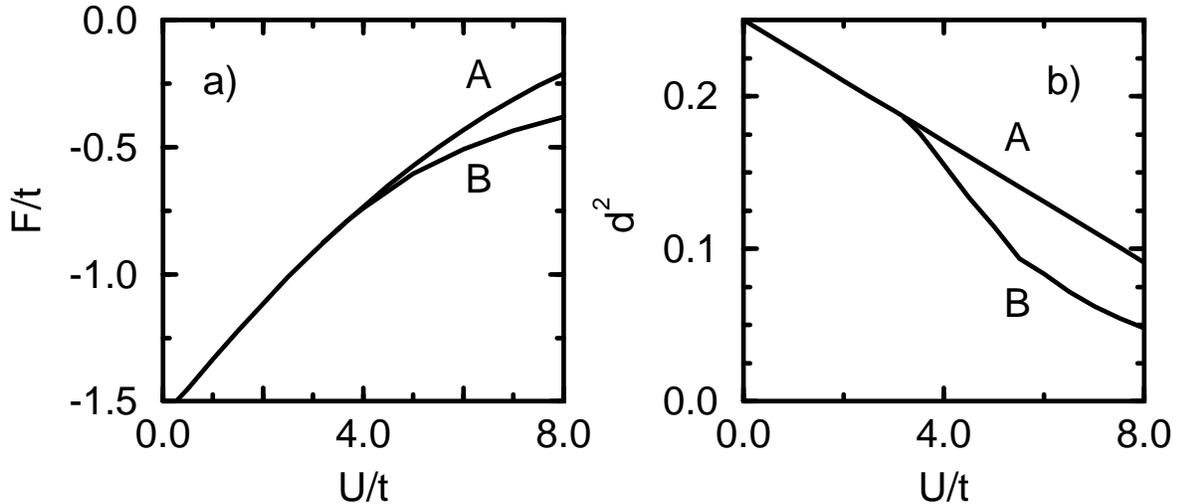

**Figure 1.** a) Ground state energy of the Hubbard Model on the honeycomb lattice at half-filling as a function of the interaction strength in the paramagnetic state (Curve A) and in the anti-ferromagnetic state (curve B).
b) Density of doubly occupied sites for the Hubbard Model on the honeycomb lattice at half-filling as a function of the interaction strength in the paramagnetic state (Curve A) and in the anti-ferromagnetic state (curve B).

In fig. 1b we show the double occupancy in both paramagnetic and anti-ferromagnetic phases. In the latter the double occupancy is seen to be reduced from its value in the paramagnetic phase in the intermediate coupling regime. In the strong coupling regime, i. e. above the Brinkman-Rice point, it will be the opposite and $d^2$ will be proportional $t/U$. As compared to the simulations [25], our results for the double-occupancy are found to be slightly smaller, especially for the largest available values of $U$. In fig. 2a we locate the metal to insulator transition point by showing the staggered magnetization $m$ as a function of $U$. It turns out that $m$ starts to deviate from 0 at $U \sim 3.1t$, a value that is fairly smaller than the one obtained in Quantum

Monte-Carlo simulations ($4.5 \pm 0.5$) but still reasonably close to it. It represents a sizeable improvement of the Hartree-Fock value (2.2). It also rules out the Brinkman-Rice scenario, as the critical $U$ is $12.6t$. The magnetic gap is given by twice the Lagrange multiplier $\beta$. It is displayed in fig. 2b. It starts to deviate from 0 at about $U/t \sim 3.1$ and grows fairly linearly with $U - U_c$ in the vicinity of the transition and is proportional to $U$ in the strong coupling limit.

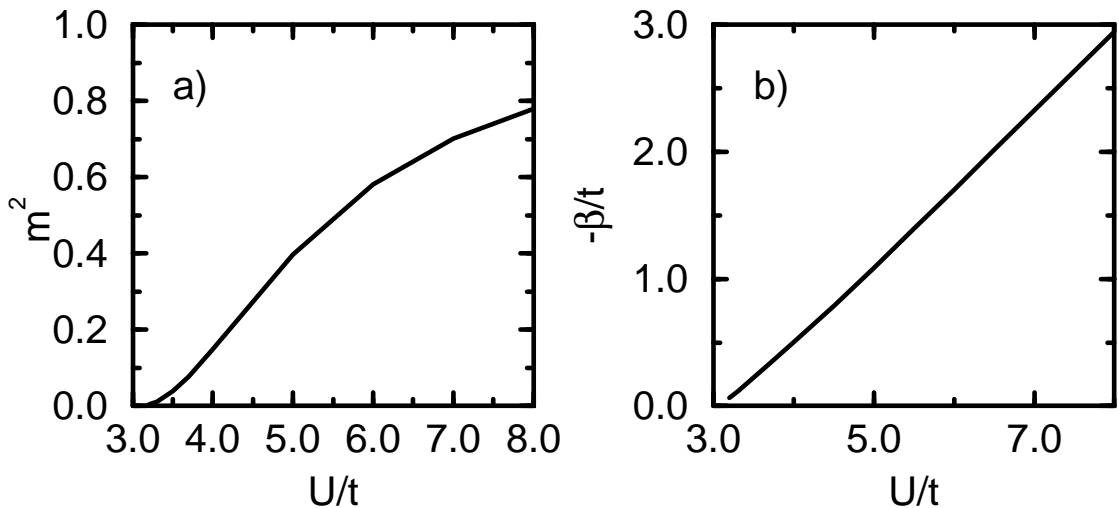

**Figure 2. a)** Square of the staggered magnetization of the Hubbard Model on the honeycomb lattice at half-filling as a function of the interaction strength.
**b)** Magnetic gap of the Hubbard Model on the honeycomb lattice at half-filling as a function of the interaction strength.

## MAGNETIC PHASE DIAGRAM

In the spin-rotation invariant slave boson mean-field theory of the Hubbard model on the square lattice the ground state mean field energy is given by:

$$F = -T \sum_{\vec{k},\sigma} \ln\left(1 + e^{-(E_{\vec{k},\sigma}/T)}\right) + Ud^2 + \alpha(e^2 + p_0^2 + d^2 - 1) - \beta_0(2d^2 + p_0^2) \qquad (7)$$

with the pseudo-fermion dispersion relation $E_{\vec{k},\sigma} = z_0^2 t_{\vec{k}} + \beta_0 - \mu_0$. Minimizing $F$ with respect to $e$, $p_0$, $d$, $\alpha$ and $\beta_0$ delivers the saddle-point. If, using the constraints, one expresses $e$ and $p_0$ as functions of $d$ and the density, one immediately obtains the free energy (7) as expressed in the Gutzwiller approximation.

As discussed by many authors, those saddle-point conditions can be cast into a single one:

$$\frac{U}{U_c} = \frac{1 - (e+d)^2}{(e+d)^4 - \delta^2}(e+d)^4 \qquad (8)$$

where

$$U_c = -8 \int_{-\infty}^{\infty} d\epsilon \, \epsilon N(\epsilon) f_F(\frac{z_0^2}{T}(\epsilon - \mu_{eff})) \qquad (9)$$

where $N(\epsilon)$ denotes the density of states per spin, and $\mu_{eff} = \frac{\beta_0 - \mu_0}{z_0^2}$. As usual $e$ and $d$ are related by $e^2 - d^2 = \delta$, $\delta$ being the hole doping. We here adopt a new attitude in front of the saddle-point condition eq. (8). Instead of solving directly eq. (8), we first choose a value for $e$ and the density, and we determine numerically $\mu_{eff}$. This being settled, we calculate $U_c$ via its definition, and we finally determine $U$ with the help of eq. (8). The result of this investigation is displayed in fig. 3. We show $z_0$ as a function of $U$ at the finite temperature $\frac{t}{6}$ for several densities.

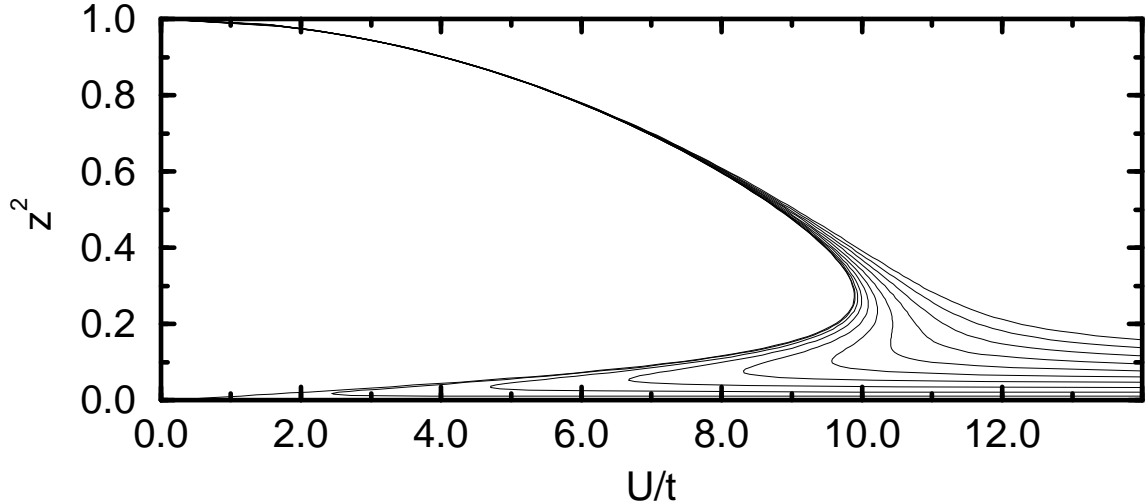

**Figure 3.** Solution of eq. (8) for several densities. They run on the right border from $\delta = 5\%$ for the upper curve down to $\delta = 0$ for the lower curve by steps of $0.5\%$.

Contrary to the zero temperature case, where our approach does not yield results differing from the standard one, we see that, for very small hole doping, we obtain several solutions. Moreover we obtain a line for a first order phase transition. The latter is fairly linear and is ranging between $0 \leq \delta \leq 0.03$ and $8.5 \leq U/t \leq 10$. In detail the latter is obtained by selecting the value of $z_0$ which is leading to the lowest free energy which, at half-filling, takes the value $-T \ln 4$. This also gives a rough estimate of the Néel temperature. At half-filling and for large $U$, the mean-field free energy in the anti-ferromagnetic state is:

$$F_{AF} = -4\frac{t^2}{U}. \qquad (10)$$

Thus long-ranged anti-ferromagnetic orders disappears for

$$T \geq (4t^2)/(U \ln 4). \qquad (11)$$

However this approach only brings a first order transition, the result (10) being essentially insensitive to the temperature. Moreover it misses the spin degrees of freedom, as $\ln(2)$ should appear in eq. (11) instead of $\ln(4)$. However the above-mentioned difficulties seem to be restricted to the close vicinity of half-filling. Indeed this approach is expected to make sense when the temperature is much smaller than the effective band width, which is not the case when $z_0$ is vanishingly small. It thus deserves a further study of the stability of the paramagnetic state with respect to an incommensurate long-ranged magnetically ordered state in the overall phase diagram. Moreover Moreo recently [29] calculated the uniform spin susceptibility $\chi_S$ in a Quantum Monte-Carlo simulation. She obtained that, for moderate interaction strength ($U/t = 4$) $\chi_S$ uniformly decreases as a function of the hole doping, whereas for strong interaction ($U/t = 10$) it first passes through a maximum, which is also seen in exact diagonalization for the $t-J$ model at $J/t = 0.4$ [30], a result that can not be obtained in RPA calculations. There would then be a crossover between small and strong coupling behavior, even though the crossover line is not yet established [31]. Let us than see what our approach can say about it.

Considering the quantum fluctuations around the saddle-point eq. (7) brings out 2 distinct channels, spin symmetric and spin antisymmetric, providing a microscopical basis for a Landau-Fermi liquid theory. It is by considering the antisymmetric channel that one obtains the spin susceptibility. In the static limit the divergence of the latter yields the phase boundary of the paramagnetic state and allows to give a magnetic phase diagram of the Hubbard model on the square lattice at finite temperature. As derived by Li, Sun and Wölfle [19], the spin susceptibility is given by:

$$\chi_S(\vec{q},\omega) = \frac{\chi_0(\vec{q},\omega)}{1 + A_{\vec{q}}\chi_0(\vec{q},\omega) + A_1\chi_1(\vec{q},\omega) + A_2(\chi_1^2(\vec{q},\omega) - \chi_0(\vec{q},\omega)\chi_2(\vec{q},\omega))} \quad (12)$$

where:

$$\chi_n(\vec{q},\omega) = -\sum_{\vec{p},\omega_n,\sigma} (t_{\vec{p}} + t_{\vec{p}+\vec{q}})^n G_0(\vec{p}, i\omega_n) G_0(\vec{p}+\vec{q}, \omega+i\omega_n) \quad n = 0,1,2 \quad (13)$$

and the others symbols are defined in ref. [19]. Setting $\omega = 0$ we obtain the instability line of the paramagnetic phase towards an incommensurate magnetically ordered state.

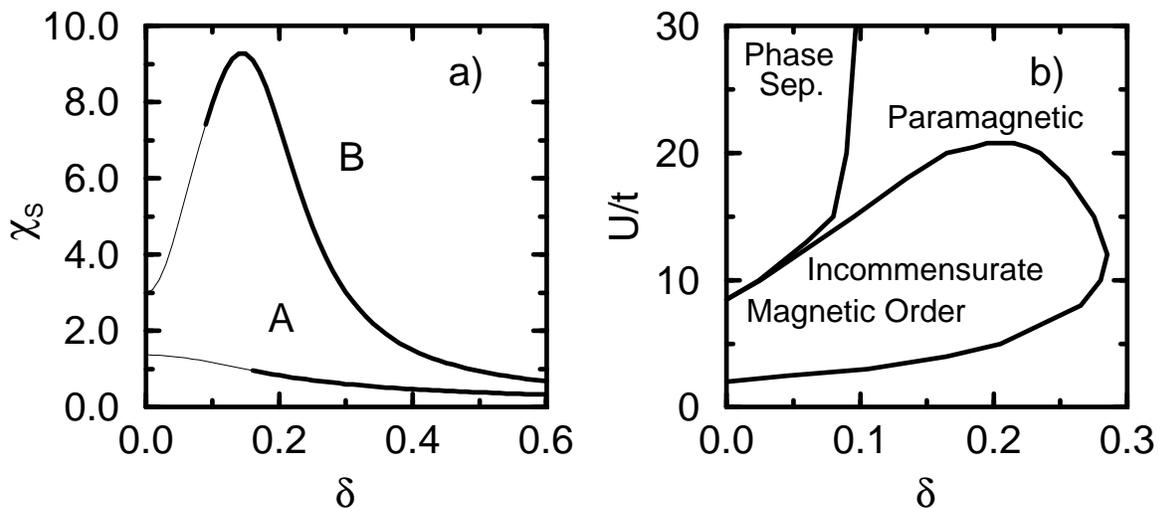

**Figure 4.** a) Static Spin Susceptibility as a function of the hole doping for $U/t = 4$ (curve A) and $U/t = 25$ (curve B) at temperature $t/6$. The thick lines indicate that the calculation has been performed in a domain where the homogeneous paramagnetic state is stable.
b) Phase diagram of the Hubbard Model on the square lattice at temperature $t/6$ as obtained from our approximation.

At very small temperature ($t/T \geq 10$) the instability line is essentially unchanged compared to the zero temperature result [26]. But there is a sudden change at about $t/T = 8$ and we display the result at $t/T = 6$. The very wide magnetic domain reduces to a small pocket around half-filling and $U/t = 10$, and otherwise the paramagnetic state remains stable. This finds an explanation in term of the magnetic gap that appears to be very small in the spiral state at zero temperature as discussed in [23, 24] in the largest part of the phase diagram. Allowing for a small temperature mixes the magnetic sub-bands and as a result the long-ranged magnetic order disappears. This is again a result that can not be obtained in RPA calculations, as there one obtains a very large magnetic gap that does not allow for a mixing of the 2 magnetic sub-bands at low temperature.

In the thus stable paramagnetic state one can then compute the uniform magnetic susceptibility $\chi_s$. The results are displayed in fig. 4a. In the limit of small $U$ the above expression for the magnetic susceptibility eq. (12) reduces to the RPA result and we obtain that $\chi_s$ is monotonically decaying upon doping. However for large $U$ $\chi_s$ first passes over a maximum before decaying as well. This maximum is reminiscent from the one that is leading to ferromagnetic instability at zero temperature [24]. The obtained behavior is in agreement with the numerical simulation of Moreo [29]. Our result implies that it is not necessary to invoke next-nearest neighbor hopping in order to obtain such a behavior. That our result for $\chi_s$ yields too large numbers is not very surprising as we are calculating in the vicinity of a magnetic instability and a better calculation is required in order to obtain a better agreement. All together our results lead to the phase diagram that is displayed in fig. 4b. It consists of 3 regions. The largest part, including half-filling for $U/t \leq 3$, and all densities for $U/t \geq 21$, is paramagnetic. This means that allowing for small temperature effects is sufficient to suppress the magnetic instability that was present at $T = 0$ [26]. It only remains at intermediate coupling and small doping. In contrast to the $T = 0$ case, a new phase separated domain appears for strong coupling and small doping. It is characterized by a negative compressibility and is paramagnetic. The 2 components entering a Maxwell construction would be paramagnetic with different densities and effective masses, one of them being infinite.

## CONCLUSION

In this paper we used the Kotliar-Ruckenstein slave boson technique to treat the strong correlation of the Hubbard model. In a first part we discussed the metal to insulator transition that is occurring on the honeycomb lattice at half-filling. We find the critical value for the interaction strength to be $U/t = 3.1$. Above it a gap opens in the spectrum which behaves like $(U - U_c)$ in the vicinity of the transition, and like $U$ for strong coupling. In a second part we calculated the phase diagram of the Hubbard model on the square lattice at finite temperature. The Mermin-Wagner theorem is fulfilled in the largest part of the phase diagram. In the strong coupling regime, the uniform magnetic susceptibility shows a maximum in its doping dependence in agreement with numerical simulations.

## ACKNOWLEDGMENTS

We gratefully acknowledge Prof. P. Wölfle for many enlightening discussions. RF thanks the Deutsche Forschungsgemeinschaft for financial support under Sonderforschungsbereich 195 .

## REFERENCES

* Present Address: Max-Planck-Institut für Physik Komplexer Systeme, Außenstelle Stuttgart, Postfach 80 06 65, D-70506 Stuttgart.